\documentclass[sigconf]{acmart}

\AtBeginDocument{%
  \providecommand\BibTeX{{%
    \normalfont B\kern-0.5em{\scshape i\kern-0.25em b}\kern-0.8em\TeX}}}

\copyrightyear{2023} 
\acmYear{2023} 
\setcopyright{acmlicensed}
\acmConference[CHI '23]{Proceedings of the 2023 CHI Conference on Human Factors in Computing Systems}{April 23--28, 2023}{Hamburg, Germany}
\acmBooktitle{Proceedings of the 2023 CHI Conference on Human Factors in Computing Systems (CHI '23), April 23--28, 2023, Hamburg, Germany}
\acmPrice{15.00}
\acmDOI{10.1145/3544548.3581513}
\acmISBN{978-1-4503-9421-5/23/04}

\usepackage{booktabs}
\usepackage{dcolumn}
\usepackage{tabularx}
\usepackage{framed}
\usepackage{graphicx}
\usepackage{tabularx}
\usepackage{graphicx}
\usepackage{url}
\usepackage{multirow}
\usepackage{enumitem}
\usepackage{hyperref}
\usepackage{amsmath}
\usepackage{lscape}
\usepackage{longtable}
\usepackage{lscape}
\usepackage{xcolor}
\usepackage{color, colortbl}
\usepackage{siunitx}
\usepackage{microtype}
\usepackage{etoolbox}
\usepackage{makecell}
\usepackage{multicol}
\usepackage{todonotes}
\usepackage{subcaption}
\usepackage{fontawesome}
\usepackage{tabularx}
\usepackage{balance}

\begin{document}

\title[Ophthalmologists' Perceptions of Anchoring Bias Mitigation in Clinical AI Support]{``If I Had All the Time in the World'': Ophthalmologists' Perceptions of Anchoring Bias Mitigation in Clinical AI Support}

\author{Anne Kathrine Petersen Bach}
\email{abach17@student.aau.dk}
\affiliation{%
  \institution{Aalborg University}
  \city{Aalborg}
  \country{Denmark}
}
\author{Trine Munch Nørgaard}
\email{tnarga16@student.aau.dk}
\affiliation{%
  \institution{Aalborg University}
  \city{Aalborg}
  \country{Denmark}
}
\author{Jens Christian Brok}
\email{jbrok17@student.aau.dk}
\affiliation{%
  \institution{Aalborg University}
  \city{Aalborg}
  \country{Denmark}
}
\author{Niels van Berkel}
\email{nielsvanberkel@cs.aau.dk}
\orcid{0000-0001-5106-7692}
\affiliation{%
  \institution{Aalborg University}
  \city{Aalborg}
  \country{Denmark}
}

\renewcommand{\shortauthors}{Bach et al.}

\begin{abstract}
Clinical needs and technological advances have resulted in increased use of Artificial Intelligence (AI) in clinical decision support. However, such support can introduce new and amplify existing cognitive biases. Through contextual inquiry and interviews, we set out to understand the use of an existing AI support system by ophthalmologists. We identified concerns regarding anchoring bias and a misunderstanding of the AI's capabilities. Following, we evaluated clinicians' perceptions of three bias mitigation strategies as integrated into their existing decision support system. While clinicians recognised the danger of anchoring bias, we identified a concern around the impact of bias mitigation on procedure time. Our participants were divided in their expectations of any positive impact on diagnostic accuracy, stemming from varying reliance on the decision support. Our results provide insights into the challenges of integrating bias mitigation into AI decision support.
\end{abstract}

\ccsdesc[500]{Human-centered computing~Human computer interaction (HCI)}
\ccsdesc[500]{Human-centered computing~Collaborative and social computing}
\ccsdesc[300]{Applied computing~Health care information systems}

\keywords{Cognitive bias, anchoring bias, decision support, artificial intelligence, AI support, bias mitigation, CDSS, DSS, ophthalmology}

\maketitle

\section{Introduction}
\begin{figure*}[t]
    \centering
    \includegraphics[width=\textwidth]{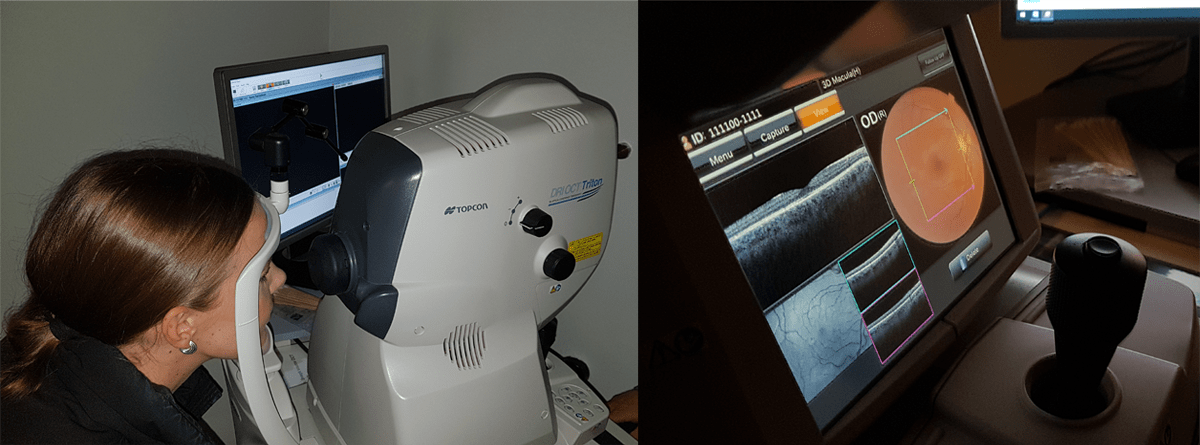}
    \caption{Imaging device in use at an ophthalmology clinic (left) and the resulting OCT and fundus images (right).}
    \Description{Two photos. On the left, a person puts their chin on a rest while looking into a machine for scanning their eyes. On the right, a computer screen shows the resulting OCT and fundus image for analysis, a joystick is visible in front of the screen.}
    \label{fig:patient_scan}
\end{figure*}

Healthcare specialists increasingly rely on Clinical Decision Support Systems (CDSS) to interpret patient symptoms and determine treatment.
In this work, we focus on Artificial Intelligence (AI)-powered CDSS in the context of ophthalmology (\textit{i.e.}, clinical care specialising in eye and vision care).
In Denmark, where our study took place, ophthalmologists experience a burdensome workload~\cite{Albinus2021Numbers}. 
Data from 2018 showed that the country's 383 ophthalmologists were responsible for close to 700.000 patients~\cite{numberseyedoctor, Albinus2021Numbers}.
To help alleviate the ophthalmologist's workload and decrease the impact of human error, AI decision support is increasingly integrated into the decision-making process~\cite{Kapoor2019AIOpthalmology}.
Specifically, the Department of Ophthalmology at Aalborg University Hospital uses an AI system to detect diabetic retinopathy (DR).
DR is an eye disease caused by high blood sugar damaging the eye's retina and can result in blindness if not diagnosed and treated timely~\cite{americanacademy}.
The detection of DR and many other eye diseases is done through visual inspection of the retinal fundus and optical coherence tomography (OCT) images.
Fundus images show the eye's interior surface, whereas OCT provides a cross-sectional image of the retina (see Figure~\ref{fig:patient_scan}).

With the increasing use of AI support systems, prior work in HCI and beyond has begun to assess the impact of such systems in real-world use.
In this study, we set out to qualitatively investigate the role of anchoring bias in clinical practice as introduced by AI support systems.
Anchoring bias is a widely established cognitive bias where a person's decisions are affected by a previously observed or experienced reference point.
Tversky \& Kahneman describe how ``\textit{people make estimates by starting from an initial value that is adjusted to yield the final answer}''~\cite{Tversky1974JudgmentUncertainty}, with any adjustments to this initial value typically being insufficient to compensate for the error of the initial reference point~\cite{Tversky1974JudgmentUncertainty}.
In clinical decision-making, anchoring bias can result in severe consequences.
While it is challenging to quantify the effect of anchoring bias on clinical outcomes precisely, numerous case studies have highlighted the real-world implications of this bias~\cite{Horowitz2021CovidAnchor, Ogdie2012ResidentReflection, Vickrey2010NeurologistsThink}.
As a timely example, Horowitz et al.\ describe how two febrile patients, seen amidst the COVID-19 pandemic, were misdiagnosed as having COVID-19 instead of their tick-borne infections due to anchoring bias induced by the many encounters with patients~\cite{Horowitz2021CovidAnchor}.
Ogdie et al.\ surveyed 41 medical residents on their prior diagnostic errors, with 88\% of respondents pointing to anchoring bias as the most common cognitive bias~\cite{Ogdie2012ResidentReflection}.
While AI-based image recognition has shown remarkable performance in diagnosis~\cite{kruse2020effects, tao2020accuracy}, adopting these systems in clinical practice often results in numerous challenges~\cite{Cabitza2019ProofPudding}.
Prior work has sought to address concerns regarding the equity and fairness of recommendations~\cite{Wang2020FactorsFairness}, the presentation of relevant and timely information~\cite{Cai2019Imperfect, Berkel2022Initial}, and alignment with clinical workflows~\cite{Cai2019HelloAI, Piorkowski2021AIDevelopersCommunicate}. 
While these are all critical aspects of designing and using AI systems, the role of cognitive biases---and in particular end-users' perceptions of techniques to combat these biases---has remained underexplored.

To better understand clinicians' use of AI-powered CDSS, we conducted an initial study of ophthalmologists' screening processes totalling seven interviews and three contextual inquiry sessions.
Specifically, we aimed to investigate how our participants use the CDSS, how it affects their screening process, and understand their motivations for (not) using the CDSS.
While the presentation of the AI's assessment (annotation of patient images) increases clinicians' efficiency, it is also perceived as deceiving as the image analysis focuses solely on DR, thereby excluding conditions such as glaucoma or tumours.
Using our newfound understanding of the clinicians' use of the CDSS, we designed an interactive prototype which integrates three bias mitigation strategies.
These established bias mitigation strategies---`\textit{hear the story first}'~\cite{Lighthall2015Understanding}, `\textit{decision justification}'~\cite{Isler2020Activating, Lambe2016Dual}, and `\textit{consider the opposite}'~\cite{Mussweiler2000Overcoming, Adame2016Training}---were designed to reduce clinicians' anchoring bias and to boost accuracy.
We evaluated our prototype with six ophthalmologists to understand their perceptions of our proposed `debiasing workflow'.

Our findings emphasise conflicting perspectives on the potential effect of bias mitigation on diagnostic accuracy, with some participants believing that their diagnostic accuracy could not be further enhanced.
Participants generally highlighted the potential of bias mitigation techniques to keep them focused throughout the workday.
However, concerns about decreasing efficiency due to these bias mitigation efforts were also common.
Participants perceived the value of not only the CDSS, but also of the bias mitigation strategies to increase with more complex cases.
This highlights not only clinicians' increased need for support in complicated cases, but also their desire to double-check their intuition.

Our results provide an understanding of clinicians' perceptions towards incorporating bias mitigation in AI-based CDSS.
We note that best practices from the bias mitigation literature can be opposite to clinicians' real-world challenges.
In particular, the time pressure clinicians face limits the appeal of mitigation strategies which require user input.
As AI systems are increasingly deployed into the real world, critically assessing the cognitive biases introduced by their use is vital.
\section{Related work}
Amidst an increase in the interest and use of AI-driven decision support tools, cognitive biases due to these systems have become an area of concern within HCI~\cite{Echterhoff2022AnchoringBiasSequential, Nourani2021AnchoringExplainable, Wang2019TheoryDrivenExplainable, Zhang2015CognitiveRemediating, Berkel2023MAP}.
Here, we discuss related work across three key topics.
First, we assess the literature on cognitive biases in clinical work, focusing on anchoring bias due to its relevance for decision support tools.
Second, we outline prior work that has aimed to mitigate either the onset or effect of cognitive biases.
Third, we present prior work on AI-supported CDSS, highlighting the challenges in deploying these systems in the real world and recent technological advances within CDSS for ophthalmology.


\subsection{Cognitive Biases in Clinical Work}
A cognitive bias is typically defined as a systematic error in our thinking~\cite{Lilienfeld2014ErrorsBiasClinical}, \textit{i.e.}, an error that is often unconsciously ingrained in our believes.
These biases are common in everyday life, and research shows no clear relation between intelligence and propensity to cognitive biases~\cite{Teovanovic2015CogBiases}.
In the clinical world, decisions are often made under time pressure and with a large degree of uncertainty, increasing the chance for cognitive biases to emerge.
As cognitive biases can result in flawed decision making, they can have a severe impact on patient health.
According to Croskerry, over a hundred different biases have been described which affect clinical decision making~\cite{Croskerry2013Mindless}.

In this paper, we focus on anchoring bias due to the contingency of this cognitive bias arising when faced with system-driven recommendations~\cite{Echterhoff2022AnchoringBiasSequential, Nourani2021AnchoringExplainable}.
Anchoring bias is described as ``\textit{the disproportionate influence on decision makers to make judgements that are biased toward an initially presented value}''~\cite{Tversky1974JudgmentUncertainty}.
This initial reference point subsequently skews our final decision outcome.
Lieder et al.\ argue that anchoring bias is a reflection of people's rational use of their limited time and cognitive resources~\cite{Lieder2018Anchoring}.
In a literature review of over 40 years of research on anchoring bias, Furnham and Boo find that this bias occurs in a wide range of domains, both as studied under controlled conditions in the laboratory and in real-world decision-making~\cite{Furnham2011LitAnchoring}.
Within the medical domain, anchoring bias has been identified as the most common cognitive bias in a survey among 41 medical residents, with a large majority of 88\% respondents having encountered this bias~\cite{Ogdie2012ResidentReflection}.
Royce et al.\ summarise case studies of misdiagnosis and state that ``\textit{errors arising from cognitive bias play a role in over 50\% of identified cases of diagnostic error in ambulatory clinics and in up to 83\% of cases involving physician-reported diagnostic errors}''.
Despite this, Croskerry highlights that decision makers are often unaware of their biases, with many clinicians under-appreciating the role of cognitive biases in diagnostic error~\cite{Croskerry2013Mindless}.
This furthermore complicates the use of debiasing strategies.

Anchoring bias is particularly relevant in the context of clinical AI support systems, as the AI recommendations act as anchors in decision-making processes~\cite{Rastogi2022CognitiveAIDecision}.
Hence, in this study, we investigate ophthalmologists' perceptions towards the inclusion of bias mitigation strategies into their AI-supported CDSS.

\subsection{Mitigating Cognitive Biases}
Given the widespread occurrence of cognitive biases~\cite{Tversky1974JudgmentUncertainty} and the effect of task context on the suitability of specific bias mitigation strategies~\cite{Berkel2023MAP}, the literature has explored a variety of methods and techniques for overcoming these biases.
The use of educational strategies for recognising and subsequently addressing cognitive biases has often been suggested, although the effect of these strategies has been inconclusive~\cite{Norman2017ErrorsClinical}.
For example, Sherbino et al.\ conducted a controlled trial to investigate the effect of cognitive forcing strategies training on medical students~\cite{Sherbino2014CognitiveForcing}.
Their results showed no reduction in diagnostic error of those having completed the training as compared to the control group.
Royce et al.\ argue that studies on the effect of teaching clinical reasoning skills to reduce cognitive biases are affected by methodological problems which affect their effectiveness~\cite{Royce2019TeachingCritical}.
In their critique, Royce et al.\ point to limitations in terms of cognitive biases studied, low levels of applicability to real-world clinical reasoning and decision making, and low levels of ecological validity~\cite{Royce2019TeachingCritical}.

Within HCI, different works have explored how changes to the interface and overall interaction may serve to mitigate cognitive biases.
Highly relevant to our work is a recent study by Fogliato et al.~\cite{Fogliato2022WhoFirst}.
In a study involving 19 radiologists, Fogliato et al.\ presented participants with an image analysis task.
Participants were shown X-ray images either simultaneously with an AI inference or were shown the AI inference after submitting their provisional diagnosis of the X-ray image.
Their results show that the radiologists who registered their provisional responses before being shown the AI inference were less likely to agree with the AI and perceived its assessments as less helpful.
Task completion time did not differ between the two conditions.
Zhang et al.\ studied two cognitive biases in the context of a decision support system~\cite{Zhang2015CognitiveRemediating}, `conservatism' and `loss aversion'.
Conservatism describes the behaviour in which we fail to adjust our beliefs when presented with new information, instead choosing to embrace the original information presented to us.
Loss aversion describes our disproportionate preference for avoiding losses over acquiring comparable gains.
With the aim of minimising these biases, Zhang et al.\ introduce two debiasing techniques; a model bootstrapped on prior suggestions and presenting the expected return for the various decision options.
Their results showed that the two studied biases have a large impact on participants' decision making, with the evaluated debiasing techniques being only partly effective.
Solomon introduces and studies `customisation bias' in the context of decision support systems~\cite{Solomon2014Customisation}.
Using a fantasy baseball game as an experimental design, participants were led to believe that they could customise the recommendations provided by the decision support system to favour specific baseball-related aspects (\textit{e.g.}, batting average).
The results of this study subsequently showed that participants who were under the impression that they had customised the recommendation system were more likely to follow its recommendations, regardless of the accuracy of these recommendations~\cite{Solomon2014Customisation}.

Cognitive biases have also been found and studied outside the context of decision support systems.
In a study on information search, Rieger et al.\ sought to mitigate confirmation bias by obfuscating attitude-confirming search results (\textit{i.e.}, results which are in line with a user's existing attitude)~\cite{Rieger2021ItemReinforce}.
Their results showed that by targeted obfuscation of moderate and extreme attitude-confirming search results through means of a warning message, participants were less likely to engage with this type of result.
However, the authors warn that this effect might also be impacted by participants' desire to avoid additional effort.
In a study on conversational agents, Santhanam et al.\ found that participants who were presented with a textual or numerical anchor prior to rating the conversation agent were more likely to provide a rating that is closer to the value of the anchor presented~\cite{Santhanam2020BiasConversational}.
This highlights that cognitive biases, such as anchoring bias, may also affect the outcomes of evaluation studies in HCI.
Recent work by Dingler et al.\ explores a method for surfacing implicit cognitive biases, in particular implicit associations~\cite{Dingler2022BIAT}.
Measuring participants' responses to (visual) stimuli, known as the implicit association test~\cite{Sriram2009BIAT}, enables researchers to measure people's leanings on polarising topics.

These collective results highlight the challenges in mitigating cognitive biases, with no `silver bullet' solution in sight.
In light of the long-term goal of mitigating cognitive biases in real-world clinical contexts, we seek to understand end-users' perceptions of bias mitigation strategies.

\subsection{AI-supported CDSS}
Recent years have seen a rise of AI-powered decision support in the medical domain~\cite{Jiang2017AIHealthcare}.
Spanning a variety of computational techniques (\textit{e.g.}, image recognition) and application areas (\textit{e.g.}, medical imaging, surgery), these applications typically aim to assist clinicians in diagnosis or treatment.
Despite promising results in terms of these system's technical capabilities~\cite{Jiang2017AIHealthcare}, real-world deployments often reveal and introduce a wide range of challenges that inhibit or limit the system's capabilities~\cite{Wang2021BrilliantAIDoc, Molin2016AutomatedImage, Lee2021CollabRehabilitation}.
For example, Wang et al.\ studied the deployment of an AI-based decision support system in a rural context~\cite{Wang2021BrilliantAIDoc}.
Their findings highlight how the high workload of clinical staff makes it practically impossible to enter the required information into the CDSS.
Wang et al.\ conclude that while the AI system might be technically competent, it is designed to support a `textbook' workflow rather than clinical reality.
Similarly, in a study on CDSS in the heart pump implant decision process, Yang et al.\ found that current decision support systems lack both clinician's trust and perceived need~\cite{yang2016investigating}.
The authors point to the CDSS's sole focus on the final decision making point, whereas decision support could prove more beneficial when presented across the patients' entire healthcare trajectory.

In a study on the integration of automation in digital pathology, Molin et al.\ study the reflections of pathologists following the introduction of a partially automated digital workstation~\cite{Molin2016AutomatedImage}.
One of their findings highlights how a lack of transparency in the inner workings of algorithms can result in end-users making incorrect assumptions about either the algorithm or the source material analysed~\cite{Molin2016AutomatedImage}.
Cai et al.\ presented an investigation in the same domain, in which they focus on the user interface used by pathologists~\cite{Cai2019Imperfect}. 
By providing pathologists with the ability to refine AI suggestions based on standard medical concepts rather than merely the visual similarity of the image, Cai et al.\ were able to increase both the perceived diagnostic utility of the tool and end-user trust.
Jo et al.\ similarly sought to improve the alignment between clinicians and a CDSS tool, focusing on the planning of discontinuation of psychiatric drugs~\cite{Jo2022LongRegimens}.
Following a user-centred approach, their study identified a need for incorporating interpersonal (\textit{e.g.}, health literacy) and infrastructural constraints (\textit{e.g.}, insurance coverage of prescriptions), aspects not previously considered in the CDSS.
These examples highlight the value of end-users' agency in collaborating with CDSSs for technology acceptance.

Finally, we highlight prior work focused on ophthalmology, in particular diabetic retinopathy.
Diagnosis in ophthalmology is based mainly on the analysis of medical images, a task that is time-consuming, strenuous, and error-prone~\cite{Kapoor2019AIOpthalmology, Nguyen2020DRDeepLearning}.
The image-centred nature of ophthalmology aligns well with the pattern recognition abilities of AI technology, with hopes of increased efficiency and quality when introduced into clinical practice~\cite{Kapoor2019AIOpthalmology}. 
Consequently, the use of AI for the analysis of patients' medical images has been explored for a variety of eye-related conditions.
Nguyen et al.\ present an automated classification system for diabetic retinopathy~\cite{Nguyen2020DRDeepLearning}, with other work focusing on retinal anomalies such as macular edema (excessive fluids in the centre of the retina)~\cite{Roy2017ReLayNet}, exudates (fluid leaking out of blood vessels)~\cite{Akram2014Macula}, and cotton-wool (localised reduced blood flow)~\cite{Niemeijer2007CottonWool}, which are typically the result of a specific condition.
Despite these technological advances, a recent review on the current state of AI in ophthalmology warns of existing limitations~\cite{Kapoor2019AIOpthalmology}, citing risks of clinicians' deskilling, AI's lack of understanding of underlying clinical contexts, and a lack of transparency of algorithms' underlying mechanisms.
Beede et al.\ evaluate the use of an AI-support CDSS for the detection of diabetic retinopathy in clinical practice~\cite{Beede2020DLDR}. 
Through interviews and observations across clinics in Thailand, the authors identified high variability in clinical practice between the clinics and a continuous organisational strain to deal with the extensive volume of patients.
Beede et al.\ subsequently studied the challenges introduced by the deployment of the AI-supported CDSS.
These focused on obtaining patient consent, working around the study protocol to better align with patients' needs, and clinical factors negatively affecting system performance. For example, the authors point to the CDSS having stringent image quality guidelines, disqualifying images as blurry which are still interpretable by clinical staff~\cite{Beede2020DLDR}.

Integrating AI support into clinical practice has remained a challenge, with a broad call for further research on how CDSSs can be successfully implemented into the clinical context~\cite{Wang2021BrilliantAIDoc, Gulshan2016DRAlgo}.
We build upon this prior work by investigating ophthalmologists' perceptions of bias mitigation as integrated into their existing AI-supported CDSS.
\section{Detecting Diabetic Retinopathy}
The Department of Ophthalmology at Aalborg University Hospital deals with all kinds of eye-related diseases and ailments.
The focus of our study is the screening of DR.
DR poses a severe threat to diabetes patients, and spans four stages of severity.
First, the illness begins with the emergence of microaneurysms in the retina.
Second, more severe changes occur in the retina, such as haemorrhages, which can affect the patient's vision.
Third, scar tissue forms, which can lead to retinal detachment---requiring immediate surgery. 
Fourth, the disease evolves into diabetic maculopathy, where repeated treatment is needed and severe vision loss may be inevitable~\cite{nhs2021StagesDiabetic}.

\begin{figure*}[t]
    \centering
    \includegraphics[width=\textwidth]{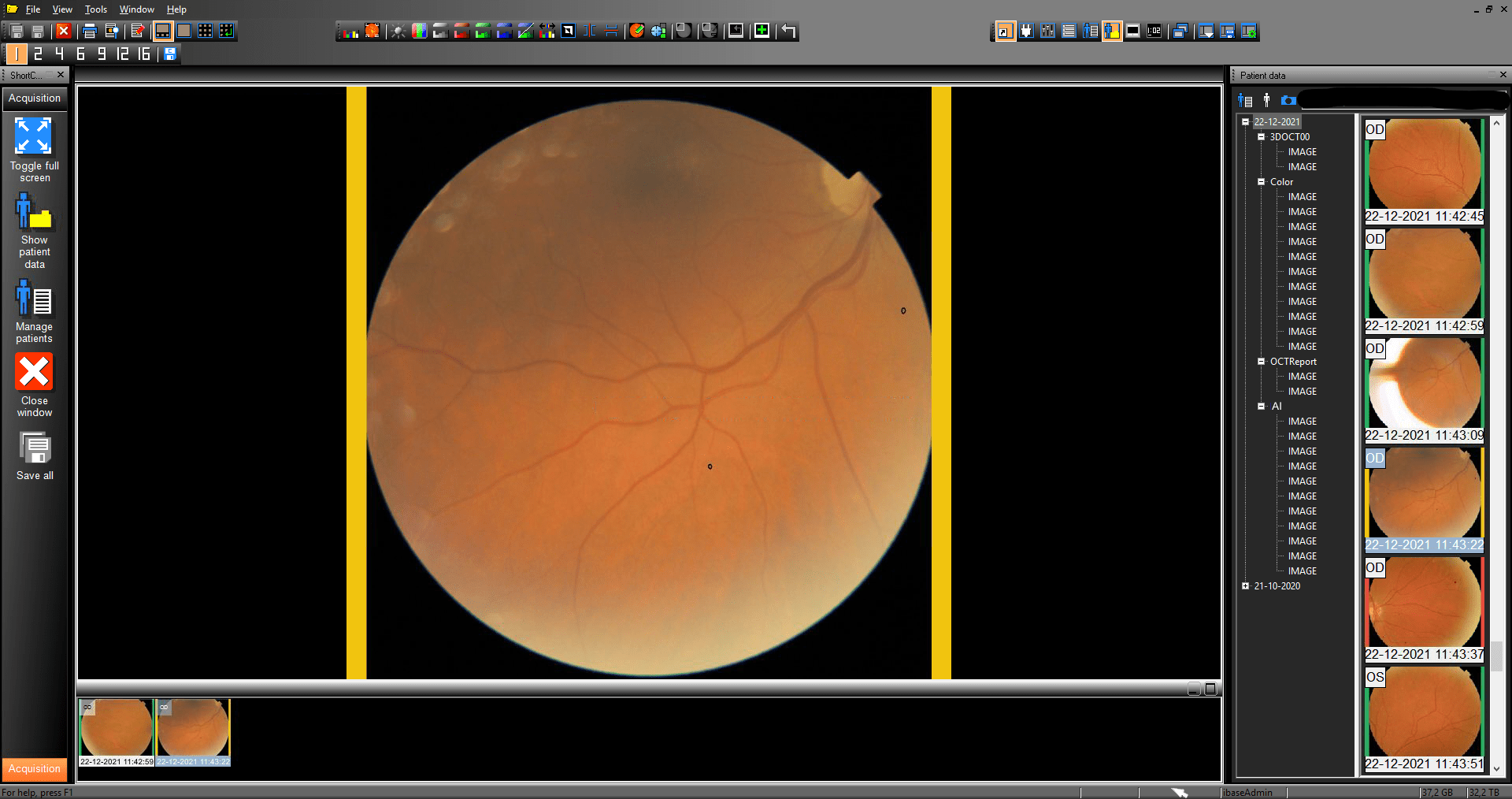}
    \caption{Assessment of fundus images as presented in the AI-powered CDSS. Yellow bars indicate that one to three lesions have been detected. Two black outlines towards the centre and right-hand side of fundus image indicate lesions.}
    \Description{Screenshot of AI-powered CDSS system. Centrally visible on the screen is a fundus image of a patient's eye. Thick yellow lines are visible on the left- and right-hand sides of the fundus image. The rest of the interface shows additional fundus images of the same patient.}
    \label{fig:CDSS}
\end{figure*}

Diabetes patients in Denmark are offered regular screenings to monitor DR-related changes in their retinas.
As these patients typically attend screenings related to other potential complications, the hospital offers patients to bundle all these appointments in one day.
During this day, patients go through various tests, including having images of their retina taken.
Instead of waiting several days or weeks to receive all test results, patients receive all results by the end of the day.
Specifically, using the equipment shown in Figure~\ref{fig:patient_scan}, a nurse or bio-analyst creates an OCT scan and takes five fundus images per eye.
The retinal images and the results from the various tests are collected and stored electronically.
The ophthalmologist assigned to screening duty for the day assesses each of the images from the same-day screening as they appear in the system.
As the patients wait at the hospital to receive their screening results, the doctor finalises each screening within one hour of receiving the images.
Here, they look through the patient's medical record, their OCT and fundus images, and add their assessment to the medical record.
Depending on the result, the patient is either called in for a consultation or scheduled for a subsequent screening.

In 2018, the hospital introduced an AI system to assist in fundus image assessment, aiming to ease the doctors' workload and increase the efficiency of same-day screenings.
This AI system is part of the screening software currently used by the doctors, and is provided by a commercial company operating across Europe.
The system analyses fundus images and can detect microaneurysms and haemorrhages, which are signs of DR.
The use of such automated systems for detecting DR lesions is increasingly common due to rising patient numbers and healthcare costs, both in clinical settings (as studied here) by opticians, and in primary care~\cite{Abramoff2018PivotalDR}.
Goh et al.\ provide an overview of commercially available systems~\cite{Goh2016RetinalScreening}, which differ in functionality between detecting DR (as studied here) or classifying cases as either `high' or `low' risk for DR and provide a subsequent referral recommendation.
The system we investigate integrates with the hospital's electronic patient record system.
When an ophthalmologist opens a patient's fundus images, they are presented with ten unprocessed images followed by the same ten images, but processed and labelled by the AI system. 
The output presented to the doctors is shown in Figure~\ref{fig:CDSS}.
The left and right sides of each image's thumbnail are coloured either green, yellow, or red, depending on the number of lesions detected.
When an image is viewed in full size, the colour label of that image lines its left and right edge.
The system generates a black outline that follows the shape of any detected lesions.

\section{Study 1 - Contextual understanding}
\label{sec:studycontextual}
Next, we elaborate on the process of our interviews and contextual inquiry, as well as our use of reflexive thematic analysis.
Our process of data gathering was guided by the following questions:

\begin{itemize}
    \item How do ophthalmologists use the AI system?
    \item Does the AI system affect ophthalmologists in their DR screening process? In which ways?
    \item To what degree do the ophthalmologists trust the output of the system?
    \item What are the ophthalmologists' motivations for using or not using the AI system?
\end{itemize}

Our sample consisted of seven ophthalmologists, all between the ages of 35--50, with their experience ranging from less than one year to 16 years (see Table~\ref{tab:studysample}).
We prepared an initial interview guide, which we refined between interviews to obtain the most valuable insights.
Each interview lasted approximately one hour, was audio recorded, and subsequently transcribed for analysis.

Three of the seven ophthalmologists also participated in our contextual inquiry (P1, P2, and P3).
Contextual inquiry enabled us to observe the doctors' use of the CDSS in their natural surroundings.
We thereby obtained a rich understanding of the factors that influence their use of the CDSS.
Throughout the three contextual inquiry sessions, we continuously wrote notes, paying attention to the participants' comments and actions.
Each session lasted approximately four hours, during which the ophthalmologists carried out their typical tasks and one of the paper's authors occasionally asked questions.
In addition, all three ophthalmologists explained their work process and showed us the tools with which they worked.

\begin{table*}[h]
\begin{tabular}{@{}lllllcc@{}}
\toprule
\textbf{Participant} & \textbf{Primary job title}             & \textbf{Sex} & \textbf{Age} & \textbf{Years of experience} & \textbf{Study 1} & \textbf{Study 2} \\ \midrule
P1                   & Chief physician               & F  & 47  & 14  & \faCheck & \faCheck \\
P2                   & Chief physician and surgeon   & F  & 50  & 16  & \faCheck & \faCheck \\
P3                   & Specialist                    & M  & 37  & 4   & \faCheck & \faCheck \\
P4                   & Surgeon                       & F  & 44  & 13  & \faCheck &  \\
P5                   & Specialist                    & M  & 35  & 3.5 & \faCheck &  \\
P6                   & Senior registrar              & M  & 35  & 0.5 & \faCheck & \faCheck \\
P7                   & Senior registrar              & M  & 44  & 3.5 & \faCheck & \faCheck \\
P8                   & Chief physician and surgeon   & M  & 40  & 12  &          & \faCheck \\
\bottomrule
\end{tabular}
\caption{Overview of the participating ophthalmologists across Study 1 and Study 2.}
\label{tab:studysample}
\end{table*}

We analysed our transcriptions and observation notes following Braun \& Clarke's approach to reflexive thematic analysis~\cite{Braun2019TA}.
Reflexive thematic analysis allows researchers to build on their existing knowledge and cultural understanding of a domain.
Precisely, we follow reflexive thematic analysis to examine which factors affect the use of an AI-enabled CDSS, focusing on the work processes and challenges of opthalmologists.
Two of the paper's authors started by separately conducting an initial analysis of two of the same interviews.
All authors reconvened to discuss and compare the identified codes and themes.
This enabled us to build a shared dictionary of knowledge and agree on the same language, creating alignment before analysing the remaining data separately.
Once the coding of all interviews and observation notes was concluded, we again discussed our results collectively to arrive at a final set of themes.

\subsection{Findings}
We next present the most relevant findings from our contextual understanding study, focusing on the following two themes: `\textit{trust}' and `\textit{system expectations}'.

\subsubsection{Trust}
A recurring theme in our findings relates to the degree of trust the ophthalmologists expressed towards the AI system and the factors influencing this trust. 
While P2 noted that the AI has a better track record than ophthalmologists in detecting microaneurysms and haemorrhages, others often pointed out mistakes by the AI.
The most common error was the marking of harmless retinal pigments as though they were microaneurysms or haemorrhages (\textit{i.e.}, false positives).
Another common error was the AI failing to detect microaneurysms or haemorrhages (\textit{i.e.}, false negatives).
Such instances prompted P3 to say that ``\textit{The AI system does not perform well enough for me to ignore the green images}''.
P1 described an occurrence in which the system was unable to detect a large tumour, with the image labelled as green.
While the system is not trained to detect tumours, this participant stated that this obstructs her from relying entirely on the system's report, as severe cases may slip past.
During observations, both P1 and P3 experienced instances of low image quality, an aspect that the AI system does not consider but which does impact its ability to analyse those images successfully.
The ophthalmologists noted these errors as a reason for distrust in the system and, to some, a reason to refrain from using the system.

As apparent in the observations and interviews, most ophthalmologists were influenced by the AI system's assessment colours.
In validating the system's assessment, ophthalmologists repeatedly referred to the colour labels of the images.
For instance, when the AI reported only one yellow image among all green images, both P1 and P2 would check whether the yellow-labelled image was assessed correctly.
In cases where all images were labelled as green, ophthalmologists spent substantially less time analysing a patient---dropping in analysis time from roughly 15 seconds to 5 seconds per image.
One participant noted that ``\textit{you can be pretty sure that nothing is there if all the images are green}'' (P5), and ``\textit{I look through all of the images, and if it} [the AI system] \textit{says they are all green, well then I go through the images slightly faster}'' (P5). Both P1 and P4 expressed a similar sentiment, with P4 specifically pointing to an increased sense of confidence when she agreed with the AI system: ``[the green labels] \textit{just give me a feeling of security}''.
P3, who had otherwise been critical of the AI system, stated that ``\textit{the colours do not matter, unless it's all green, in which case I go through them quickly}''.
Interestingly, when asked about the meaning of each of the three colours, this participant was unsure what determined each label.
Similarly, P7 was unsure about the meaning of the yellow label.
When it came to images labelled as yellow or red, both P4 and P6 stated that they paid equally more attention to those compared to the ones labelled green.
P4 stated that ``\textit{If I have some that are yellow or red---and it really doesn't matter whether they are one or the other---then I look at them very carefully}''.

\subsubsection{System Expectations}
Our participants had varying ideas of what to expect from the system.
While some saw value in the system's abilities, others were more negative and compared it with more advanced AI systems.
Four participants (P1, P2, P4, P6) valued the system enough to regularly use it in their assessment, whereas the remaining three ophthalmologists (P3, P5, P7) found its value proposition to be too elementary.
When assessing fundus images, the doctors look for several abnormalities.
These abnormalities include cotton-wool, neovascularization, macular edema, exudates, and microaneurysms.
The AI system in use is limited to detecting microaneurysms and haemorrhages, prompting one of our participants to state ``\textit{it only finds red dots, and I can't really use that for anything}'' (P7).
Although the overall consensus was that the system should ideally detect all abnormalities in the retina, some ophthalmologists highlighted that they also valued the system's current functionality.
P1, P4, and P6 stated that the system provides a quick overview of a patient's progression of DR, as they could see the colour-based labelling on the thumbnails.
Additionally, P2, P4, and P5 explained that they make use of the colour labelling as a way of self-checking, obtaining a sense of comfort in their analysis when the system's output corresponded to their own assessment.

The AI system as used by our participants in the clinic is also used by optometrists in non-clinical settings.
The optometrists typically refer patients to the hospital in case any lesions or deviations are detected.
Given the aforementioned concerns regarding the accuracy of the system, some of our participants expressed concern regarding the use of this system by those without extensive ophthalmological training.
As optometrists may not have the necessary experience in assessing the validity of the AI's assessments, our participants expressed the perceived risk of both optometrists and patients obtaining a false sense of security if the system did not mark any lesions.
Our participants did not express a similar concern for their own assessments, even though our interviews did identify a clear pattern in the handling of `green' cases in comparison with `yellow' and `red'-labelled images.

Throughout our interviews and observations, some ophthalmologists compared the system to other AI systems and their capabilities.
For instance, P5 mentioned that it would be valuable if they could train the system by giving it feedback on its classifications, making it ``\textit{evolve over time and not be stuck}''.
P3 and P7 furthermore brought up the idea of connecting the AI system to the patient's medical record, enabling it to provide a more precise diagnosis based on a wider range of patient health details.
Besides fundus images, the ophthalmologists also assess OCT scans, which they would also want the AI system to be able to process.
\section{Study 2 - Bias mitigation}
As observed in Study~1, the AI-generated colour labels act as an anchor in the clinical decision-making process.
Our study participants expressed a change in their inspection as dependent on the label presented, a behaviour which we were able to confirm through our observations.
Further, our participants did not point to any measures taken to counter this potential bias in their clinical assessment.
Our observation of anchoring bias emerging due to the introduction of decision support aligns with recent literature~\cite{Rastogi2022CognitiveAIDecision, Fogliato2022WhoFirst}.
Prior work furthermore indicates that clinicians are unaware of the scale of cognitive biases and their effect on clinical decision-making~\cite {Croskerry2013Mindless}, which explains why no effort had been taken to address the impact of anchoring bias in the presented case.
Given the importance of minimising anchoring bias in clinical decision making~\cite{Fogliato2022WhoFirst, Furnham2011LitAnchoring, Croskerry2013Mindless}, we set out to design and assess bias mitigation strategies in the context of an existing ophthalmology CDSS.

\begin{figure*}[t]
    \centering
    \includegraphics[width=\textwidth]{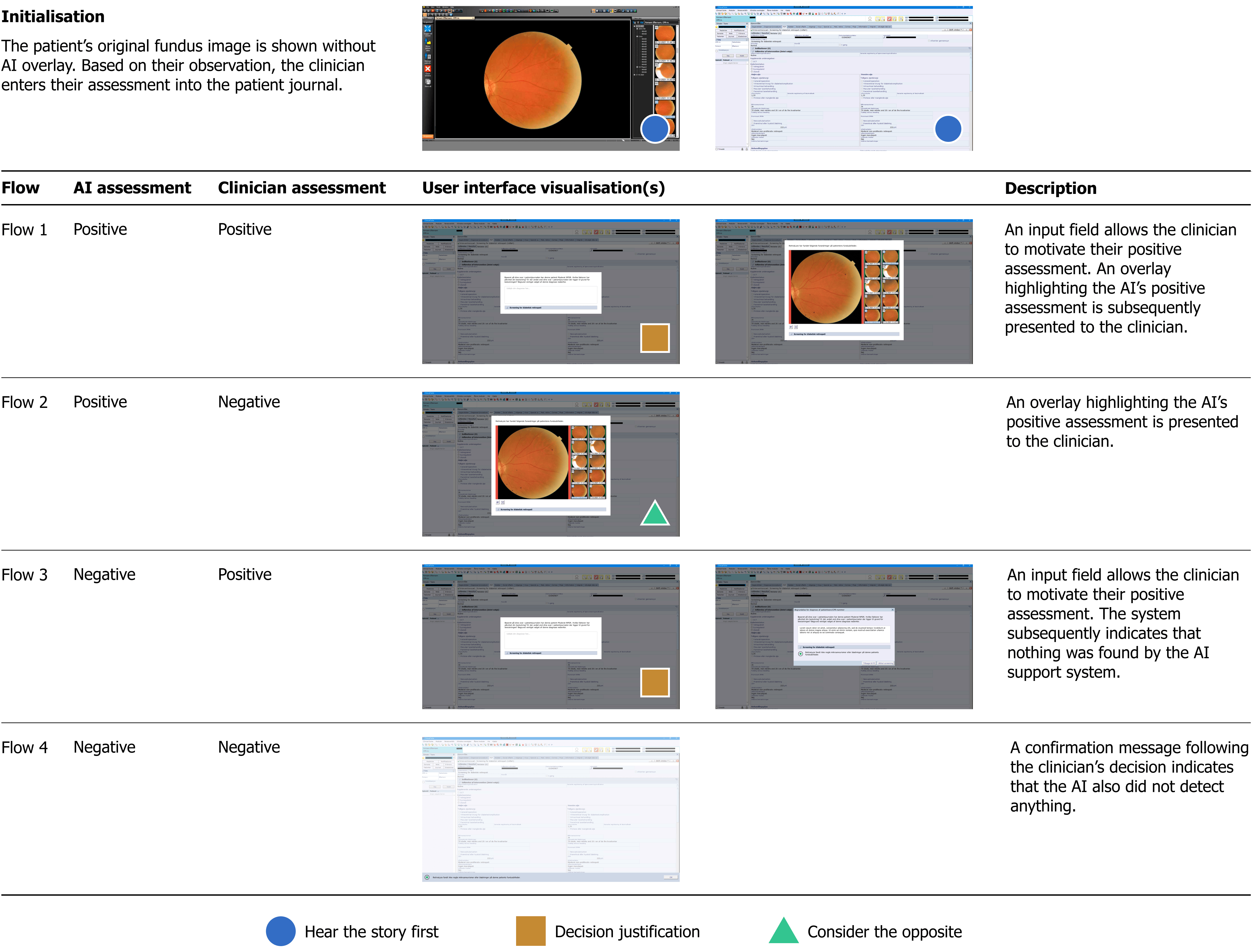}
    \caption{Overview of the presented interaction flows. Interaction flow depends both on the clinician's and the AI's assessment.}
    \Description{Overview of the workflows. The top of the image shows the current workflow, minus the AI annotation. These images are labelled as `Hear the story first'. The remainder of the image is divided into four 'flows', the presentation of which is dependent on the combination of the AI's assessment (positive/negative) and the clinician's assessment (positive/negative). In case of a positive assessment by the clinician, a popup is shown to ask the clinician for further motivation (labelled: decision justification). In case of a negative assessment by the clinician but a positive assessment by the AI, the AI's classification is shown (labelled: consider the opposite).}
    \label{fig:flowchart}
\end{figure*}

\subsection{Prototyping bias mitigation strategies}
Anchoring bias is the result of the cognitive tendency to make assessments based on the initially observed information rather than considering all subsequently received information, with the final decision largely affected by the initial impression~\cite{Lighthall2015Understanding}.
We explore three strategies for mitigating anchoring bias in the context of CDSS: `\textit{hear the story first}', `\textit{decision justification}', and `\textit{consider the opposite}'.

\subsubsection{Hear the story first}
One approach to mitigating anchoring bias is to assess the facts before being presented with the assessment of others~\cite{Lighthall2015Understanding}.
In Groopman's investigation on the thought processes that underlie medical decisions, an interviewed clinician stressed the value of avoiding anchors; ``\textit{When a case first arrives, I don't want to hear anyone else's diagnosis. I look at the primary data}''~\cite{groopman2007doctors}. 
`Hear the story first' directly addresses the observation from Study~1 that opthalmologists spent less time on the images labelled as `green' by the CDSS, even though our participants were aware that the AI's assessment is based on a limited number of clinical symptoms.
This mitigation strategy was implemented by Fogliato et al., who refer to this as a `two-step workflow', in which the AI assessment only shows after the participant has made an initial decision~\cite{Fogliato2022WhoFirst}.

In implementing this bias mitigation strategy, the clinicians can only view the original (\textit{i.e.}, not processed by the AI system) fundus images.
After looking at all the relevant data, the clinicians return to the patient journal to fill in the diagnosis, as depicted in Figure~\ref{fig:flowchart}.
Depending on the assessment of the clinician and the behind-the-scene AI assessment, the application initiates one of four pathways.
`Hear the story first' is implemented by removing any AI assessments from the initial view, as shown in the initialisation stage in Figure~\ref{fig:flowchart}.

\subsubsection{Decision justification}
This strategy prompts decision-makers to explain their reasoning to activate reflective thinking.
Doing so increases the quality of diagnosis and mitigates cognitive biases~\cite{schmidt2014exposure}.
Further, `decision justification' was identified as a consistently successful strategy in improving diagnostic accuracy~\cite{Isler2020Activating, Lambe2016Dual}. 

Through our observations in Study~1, we learned that ophthalmologists already describe any positive diagnosis in the hospital's electronic patient record following their assessment.
However, the current workflow does not support active reflection, as the goal of the logging activity is documentation rather than reflection.
We, therefore, implement the `decision justification' strategy to occur prior to the presentation of the AI's assessment and whenever the ophthalmologist has identified an abnormality in the patient's fundus images (Figure~\ref{fig:flowchart}, Flow 1 and 3).
The interface presents the user with a pop-up window in which they write their motivation for this assessment, thereby activating their analytical mindset.
If the AI system has also identified an abnormality, it displays its assessment to allow the user to compare their diagnosis (Figure~\ref{fig:flowchart}, Flow 1). This is particularly relevant if the ophthalmologist and the AI system have identified different abnormalities.
If the AI system has not identified an abnormality, this information is shared with the user following their decision justification (Figure~\ref{fig:flowchart}, Flow 3).

\subsubsection{Consider the opposite}
Considering possible alternatives before deciding is a bias mitigation strategy that has been proven effective at mitigating anchoring bias~\cite{Mussweiler2000Overcoming, Adame2016Training}.
This strategy prompts the decision-maker to consider information that contradicts their current beliefs~\cite{Mussweiler2000Overcoming}.
Simply presenting decision-makers with multiple alternative options to consider has been shown to reduce cognitive biases in general~\cite{Adame2016Training}.

This bias mitigation strategy is activated whenever the ophthalmologist finds no abnormalities, but the AI system does. In this case, the user is presented with an opposing answer from the AI system once they have entered their diagnosis into the patient journal (Figure~\ref{fig:flowchart}, Flow 2). This lets the ophthalmologist consider whether or not they have missed anything in their diagnosis.

In case neither the ophthalmologist nor the AI system found DR-related abnormalities, the ophthalmologist is presented with a banner at the bottom of the screen stating that the AI system did not find any microaneurysms or haemorrhages (Figure~\ref{fig:flowchart}, Flow 4). This design choice was based on Study~1, in which we discovered that the ophthalmologists experienced a sense of comfort when their negative assessment matched that of the AI system.

\subsubsection{Integrating bias mitigation strategies}
We chose to integrate all three bias mitigation strategies in one interactive prototype.
While this prevents us from contrasting the strategies' individual effects (\textit{e.g.}, each workflow starts with the `hear the story first' strategy, possibly followed by an additional strategy---see Figure~\ref{fig:flowchart}), it ensures that we can discuss all strategies with our participant sample and consider all possible combinations of assessment outcomes.
The prototype's user interface closely resembles the application currently used in the clinic.
This guaranteed participants' understanding of the system and ensured that participants' focus was not diverted toward the visual elements of the system.

\begin{table*}[]
\begin{tabularx}{\textwidth}{@{}lXXX@{}}
\toprule
& \textbf{Hear the story first} & \textbf{Decision justification} & \textbf{Consider the opposite} \\ \midrule
Positive reflections 
& Allows clinicians to provide their own assessment without external influence.
& Possible value in organising thoughts and sharing these with other team members.
& The short interruption to the workflow allows the clinician to reconsider their initial assessment. \\
Negative reflections 
& Additional time spent analysing each image, efficiency brought in by AI system is lost.
& Additional effort required by clinicians to write down their motivation for assessment. Perceived as repetitive to the assessment already provided in the patient journal.
& The AI suggestions can be perceived as correcting the clinician rather than supporting them in their task.\\ \bottomrule
\end{tabularx}
\caption{Summary overview of participants' reflections on the three bias mitigation strategies.}
\label{tbl:summaryfeedback}
\end{table*}

\subsection{Method}
To investigate clinicians' perceptions of the bias mitigation techniques, we conducted evaluations with members of Aalborg University Hospital's Department of Ophthalmology.
Through our liaison within the department, we were able to recruit six ophthalmologists, five of which also participated in Study~1. We report participant details in Table~\ref{tab:studysample}. 
We conducted a pilot evaluation with our liaison at the department before the evaluations to identify potential errors.
All six evaluations were audio-recorded and transcribed.
Each of these evaluations, including the interview, lasted for approximately one hour.
The evaluations took place in the participant's place of work, where they have access to and regularly interact with the AI system outlined in Section~\ref{sec:studycontextual}.
Three of the paper's authors were present at all evaluations.
This allowed us to distribute tasks, including the introduction of the prototype, the interview, and taking notes.
During the evaluations, participants were presented with the prototype and walked through each of the four flows.
We subsequently asked for participants' reflections on the debiasing techniques and cognitive biases.

Following the evaluations, all transcripts were analysed using reflexive thematic analysis, following the same procedure as in Study~1.
The same authors present at the evaluations took part in analysing the transcripts.
Initially, one transcript was chosen to be analysed individually by all.
Then, all authors compared the resulting codes and themes, sharing our understandings to establish alignment.
The remaining transcripts were then divided to be analysed individually.
When all six transcripts had been analysed, we collaboratively assessed all themes and codes.

\subsection{Findings}
We identified four main themes from our evaluation results. These themes focus on the envisioned effect on diagnostic accuracy, operational efficiency, openness to AI-generated insights, and the role of bias in daily ophthalmological practice.
We summarise participants reflections on the three bias mitigation strategies in Table~\ref{tbl:summaryfeedback}.

\subsubsection{Envisioned Effect on Diagnostic Accuracy}
Diagnostic accuracy was a core discussion point in evaluating the bias mitigation techniques.
Several ophthalmologists (P1, P3, P7) could not imagine that the debiasing techniques would heighten their diagnostic accuracy, as they generally did not think their diagnoses could get any more accurate.
The rest (P2, P6, P8) firmly believed that debiasing would improve diagnostic accuracy, but at the cost of efficiency.
For instance, P6 stated: ``\textit{I think it would be really great to have periodically} [...] \textit{to get back into it and to start thinking for yourself, and not fall asleep completely}''.
This points to the fact that screening for several hours can exhaust clinicians' ability to stay focused. 
When asked to imagine that they had as much time as they wanted to conduct a screening, disregarding overall efficiency, P2 stated: ``\textit{If I had all the time in the world, this would be the optimal way to do it}''.

In relation to the implementation of `hear the story first', P2 stated that removing the AI output from the fundus image would likely increase diagnostic accuracy.
Specifically, she stated: ``\textit{The bias has been removed, I arrive at my own conclusion, and then I compare my conclusion with the AI's conclusion}'' (P2).
Four of the ophthalmologists (P2, P6, P7, P8) stated that `consider the opposite' would be the most helpful in ensuring that they do not miss any critical details.
For instance, P2 stated: ``\textit{Sometimes you get tired, so if the program said `Something is there', and I scale up} [the image] \textit{and then I think `Yes, I didn't notice that myself'}''. 
Though more sceptical, P8 stated, when referring to `consider the opposite': ``\textit{It is absolutely a more useful solution. But it wouldn't make our work any faster, but from a quality perspective, then yes}''.
Concerning the strategy of `decision justification', P6 hypothesised that by sharing one's justifications with the team, the diagnostic accuracy could be heightened collectively, as they would gain insight into each other's thought processes.

All ophthalmologists expressed that the debiasing techniques could be a learning tool for inexperienced screeners to heighten their diagnostic accuracy.
As inexperienced ophthalmologists can be less confident of their decisions, the system could help them develop their diagnostic reasoning.
For instance, P7 commented that it could ``\textit{act as a kind of safety measure}''.
In that regard, P2 stated: ``\textit{When you are training, you have a need for someone else to check what you are doing, `Is this right or is it wrong?'}''.
Similarly, P8 stated: ``\textit{It would work well as a learning tool since you have to reflect on what you have found---but to those of us that are as obdurate as we are, I don't think it will change very much}''. 

\subsubsection{The Restraining Role of Efficiency}
The clinicians' focus on efficiency in their work plays as large a role as their diagnostic accuracy.
As P1 stated, ``\textit{in our everyday work, we have to focus on efficiency mostly, we have to be quick and view many images because the patient numbers keep growing. So we need a solution that helps us become quick}''. 
Similarly, P2 reflected; ``\textit{efficiency is king today, and that is what has become problematic. Often you consciously compromise}'', indicating that the desired efficiency comes at the expense of diagnostic accuracy.
P2 added that the number of tests performed on each patient has increased, resulting in more data for each patient.
Consequently, the ophthalmologists were critical of the impact on efficiency that bias mitigation would have.
One of the aspects that exemplified this standpoint was the implementation of `hear the story first', in which the AI-generated colour labels were removed.
Specifically, two of the ophthalmologists (P1, P2) commented that they would have to spend more time examining each fundus image themselves, as opposed to getting a quick overview of all fundus images as provided by the AI colour labels.
In this regard, P2 stated that ``\textit{if I already trust what the AI system says and I have its answer from the start then I can shortcut looking through all the images. I can't do that if I get them} [the processed images] \textit{afterwards. Then I have to analyse every image as new}''.
P1 elaborated: ``\textit{There are very many patients that have nothing on their retinas}'', referring to how she can quickly identify the cases with no DR due to the AI system labelling all of these patients' fundus images green.

Other ophthalmologists (P3, P6, P7, P8) expressed indifference to removing the colour labels.
This was primarily because they did not use the AI system's output at all or did not look at it until they had gone through the original images themselves.
P7 saw potential in using `hear the story first' to avoid being biased at the beginning of screenings and elaborated that he hopes his colleagues using the colour labels do not take them too literally.
When presented with the first and third flow (Figure~\ref{fig:flowchart}), in which `decision justification' is implemented, participants again expressed concern regarding efficiency.
For instance, P8 considered it counter-productive to justify whenever he found any retinal abnormalities: ``\textit{from an efficiency point of view---no thanks. It would just prolong the process completely}''.

\subsubsection{Openness to AI-generated Insights}
Our incorporation of the bias mitigation strategies sparked reflections on participants' role as domain experts. 
As stated by P2: ``\textit{When you're an expert, you have many years of experience, but you also think you do it better than others. And then it is up to me how much doubt I allow in myself -- whether I accept other information that contradicts me}''.
This participant then suggested that some experts are more open to input than others due to their belief that they outperform automated solutions.
Most of the ophthalmologists (P2, P3, P6, P7, P8) brought up that the presentation made it seem like a precautionary measure to fact-check their answer, which pertains specifically to the strategy of `consider the opposite'. 
As P2 commented: ``\textit{At the moment, it} [the AI system] \textit{suggests `this could be a haemorrhage' and then I go onto the image and see if that is correct---so in that way, I fact-check the program. In this way, the program is the one fact-checking me}''.

Some ophthalmologists (P2, P6, P7) found the prospect of being fact-checked by the AI system appealing and an excellent way to take advantage of its capabilities.
In addition, P6 elaborated that he would use the AI system's output to go ``\textit{back to see if what I found is just nonsense}'' and that ``\textit{even though we might agree, there could still be anomalies that you would notice after that you have missed}''.
On the contrary, other participants (P3, P8) were more sceptical of the debiasing techniques due to their preexisting distrust of the AI system due to it mistaking harmless retinal pigments for DR-related abnormalities.
However, had the AI system been more capable (\textit{i.e.}, detecting more types of abnormalities and higher precision), they would be more open to using the debiasing techniques.
As P3 stated, ``\textit{if the AI was very skilled and functioning, then I could easily imagine it as some sort of mentor that would hit people over the hand and say, `Let's go through 20 images, is anything there or not?'}''. 
He elaborated: ``\textit{It does not work if it's the janitor that comes in and lectures the executive. That is kind of what is happening here. It has to be another executive, capacity-wise, someone who is capable and who knows what can be done, that comes in and lectures the executive}'', alluding to the fact that he does not recognise the AI system as able to correct his diagnosis.
This point of view is supported by P3: ``\textit{it is very rare that it} [the AI system] \textit{shows me something that will make me change my mind}''.
Contrarily, as is evident in the following statement, P2 thinks more highly of the AI system: ``\textit{I think this} [decision justification] \textit{works well because there is a disagreement between two systems, where I believe in myself, but I also believe in the system, and then it is important to document why I disagree}''.

\subsubsection{Bias in Current Practice}
The ophthalmologists recalled bias as being only a minor part of their education.
Bias was not perceived as an aspect they put a lot of thought into during their daily practice.
When asked about the role of bias in their work, P1, P3, and P6 stated that bias does not affect their work much.
Others (P2, P7, P8) thought it played a more prominent and unavoidable role in their work.
P2 explained how she sometimes sees bias as playing in her favour: ``\textit{If I want to be able to use 15 minutes looking at the patient, then I need to have a preliminary answer. That means that I want to have bias because it helps me with my own decision-making}''.
P7 explained how they work based on a \textit{working diagnosis} (\textit{i.e.}, the most probable diagnosis that the clinicians believe to be the cause of a patient's symptoms), guiding the choice of diagnostic tests to confirm or deny the working diagnosis.
As new information is discovered, the clinicians adjust by choosing among other potential causes, or \textit{differential diagnoses}, before ultimately arriving at a final diagnosis.
P2 added: ``\textit{you have to create a working diagnosis: `What do I think it could be?', `How can I treat it?', `What if it's wrong?', `What kind of mistakes might I produce from the treatment?'}''. 
P7 stated: ``\textit{You receive a diagnosis that you work from, and at this point, it is important to `reset' and instead create your own diagnosis, as it} [the symptoms] \textit{could be caused by something else. We have to provide evidence for the diagnoses we make}''.
Thus, to avoid premature closure of cases, ophthalmologists use differential and working diagnoses to ensure that alternative diagnoses are considered.

\section{Discussion}
Having collected domain experts' perceptions on the integration of bias mitigation strategies, we discuss clinicians' perspectives on the role of bias and bias mitigation in decision support systems.

\subsection{Designing for Bias Mitigation}
We have evaluated three bias mitigation strategies as integrated into an interactive CDSS prototype.
Although mitigation strategies are frequently discussed in the literature~\cite{Isler2020Activating, Lambe2016Dual, Adame2016Training, Mussweiler2000Overcoming}, relatively few studies have explored how bias mitigation can be incorporated into AI decision-support systems.
We evaluated three bias mitigation strategies, namely `hear the story first', `decision justification', and `consider the opposite', which we selected based on prior work's suggestion that these are well suited for combatting anchoring bias.
We first summarise participants' perceptions of these strategies and subsequently outline three recommendations and opportunities for future bias mitigation tools in the context of Human-AI interaction.

Our findings indicate that the ophthalmologists using AI-generated colour labels to obtain an overview of a patient's status did not favour the `hear the story first' strategy.
Those who did not use this overview expressed indifference to removing the colour labels.
All ophthalmologists expressed efficiency concerns concerning the `decision justification' strategy, reflecting the additional effort it requires by clinicians.
The ophthalmologists largely agreed that `consider the opposite' would be the most helpful strategy to avoid missing important details.
Despite this, several participants commented that they would experience such a system as if being `fact-checked' by the AI system.
These results highlight the inherent challenges in incorporating bias mitigation strategies in AI-supported CDSS.

\textbf{Recommendation 1: Introduce subtle design patterns to overcome aversion to bias mitigation interfaces}.
Given these complexities and trade-offs, we call on the HCI community to consider and evaluate alternative interface designs that could support bias mitigation strategies.
In particular, more subtle designs could help to make end-users feel less fact-checked. 
How to achieve subtleness in interactions is a relatively unexplored area within HCI~\cite{Pohl2019Subtle}, although a number of works have evaluated how to `nudge' people's attention or behaviour through interface adjustments~\cite{Yang2019Unremarkable}.
For example, Echterhoff et al.\ show that manipulating the order in which decisions are presented can reduce sequential dependencies to prior decisions, thereby reducing anchoring bias~\cite{Echterhoff2022AnchoringBiasSequential}.
Another example of subtle interaction is presented by Sridharan et al., who show that the gaze of novice radiologists can be manipulated to follow the scanpath of an expert radiologist, resulting in increased diagnostic accuracy~\cite{Sridharan2012GazeManipulation}.

\textbf{Recommendation 2: Activate bias mitigation techniques periodically rather than constantly}.
We encountered several participants that questioned the need for bias mitigation in their work. 
Prior work has highlighted that decision-makers can become overconfident in their abilities~\cite{Mussweiler2000Overcoming, Lighthall2015Understanding}.
Those questioning the need for bias mitigation described the AI system as a junior colleague presenting their opinion.
This exacerbated the notion of feeling fact-checked by the recommendation system.
To reduce the feeling of being constantly evaluated, a possible solution is to periodically activate the bias mitigation strategies.
This follows from participants' comments that bias is merely a minor part of their education and is not of serious concern to them throughout the working day.
While evidence of long-term retention effects of bias mitigation techniques is scarce~\cite{Korteling2021RetentionLit}, several studies have found positive effects of bias mitigation in studies lasting up to eight weeks~\cite{Korteling2021RetentionLit, Veinott2013Decision, Morewedge2015Debiasing}.
Such a period would suffice in the clinical context studied in this paper, thereby supporting the real-world feasibility of this recommendation.

\textbf{Recommendation 3: Increase user awareness of potential cognitive biases through the decision-support interface.}
Cognitive biases are ubiquitous, and clinical work is no exception~\cite{Croskerry2013Mindless}.
An essential step towards addressing these biases is by providing clinicians with the tools to increase and subsequently maintain their awareness of these biases.
While the clinicians in our study sample expressed concern towards the potentially detrimental effects of cognitive biases, they also pointed to a lack of coverage of cognitive biases in their formal training.
Although the need for adequate bias mitigation is increasingly recognised in both the academic literature and national guidelines for clinical training~\cite{Sullivan2018Cogbias}, reflection on the role of decision-support systems in these efforts remains rudimentary.
The interfaces with which clinicians interact throughout the day could, therefore, provide a gateway to increase awareness of cognitive biases.
Prior work has, for example, augmented the interface of a social media platform with visual hints to indicate a wider diversity of opinions to the user~\cite{Gao2018BurstBubble}.
Awareness of cognitive biases is especially critical in light of a recent literature review which indicates that cognitive biases can aggravate with the introduction of explainable AI~\cite{Bertrand2022CognitiveBiasReview}.

\subsection{Expectations and Realities of Real-World AI Adoption}
The efficiency of clinicians is of critical importance to the Department of Ophthalmology.
As the number of tests performed per patient has gone up~\cite{Albinus2021Numbers}, so has the amount of data that needs to be interpreted by doctors.
In addition, a political decision was made in Denmark to screen all diabetes patients regularly to monitor progressing DR symptoms.
This has increased the pressure on the public health care system, with some of our participants highlighting how other matters are at times neglected to ensure efficiency. 
The promise of an AI system which supports clinicians in dealing with an increasing number of cases was therefore well received by the department.

Our results challenge some of the perceived benefits of deploying an AI decision-support system into ophthalmology practice.
When discussing the implementation of bias mitigation strategies with the ophthalmologists, several mentioned that they would be more likely to favour the proposed debiasing techniques if the AI system was more capable.
Specifically, they wished for the AI system to produce fewer false positive alerts and be able to detect more types of abnormalities.
This points to a mismatch between clinicians' mental model and expectations of the AI's capabilities~\cite{Berkel2023MAP}.
Similar observations were made by Yang et al.\ in their investigation of decision-making practices in the context of heart pump implants~\cite{yang2016investigating}.
Here, the AI's quantitative predictions did not align with the information needs of the clinicians.
In our study, several participants pointed to past experiences in which the AI system missed a large tumour.
While the deployed AI system was not designed to recognise and alert to such occurrences, these `failures' to detect obvious abnormalities do negatively impact clinicians' experiences.
As detection technology advances, such problems may eventually be overcome. However, the expected timeline for this remains uncertain.
A recent literature review on AI for DR screening states that ``\textit{most [commercial] systems are still being actively developed with changes and improvements to detection algorithms, user interface, scalability, better detection of DMO, etc. in progress.}''~\cite{Grzybowski2020ReviewAIDR}.
As such, HCI research must do more to align end-users' expectations with AI systems' capabilities and ensure integration in existing work practices, as highlighted in prior work~\cite{Berkel2021VisualMarkers, Yang2019Unremarkable}.
Grzybowski et al.\ state that even if AI assessment systems increase in quality, clinical experts will still be required to assess atypical situations and low-quality images, as well as for quality monitoring~\cite{Grzybowski2020ReviewAIDR}.

In addition to the technological limitations highlighted by our participants, our results further point to the tension that may arise between the need for respecting clinicians' expertise and the need for mitigating cognitive biases in CDSS.
Prior work on AI decision support has shown that clinical experts may choose not to use such support systems due to a lack of perceived need~\cite{yang2016investigating, Wang2021BrilliantAIDoc}, especially among senior clinicians~\cite{yang2016investigating, Berkel2021VisualMarkers}.
In contrast, enabling the clinicians to direct the CDSS rather than merely receiving its recommendations has improved clinicians' acceptance and efficiency~\cite{Jo2022LongRegimens, Cai2019Imperfect, Jo2022LongRegimens}.
Bias mitigation may risk compromising clinicians' agency by intervening in their workflow, a critique also reflected in our findings.
Given the impact of clinicians' agency on the eventual adoption of CDSS, this topic warrants critical consideration in future work on bias mitigation.

Lastly, while the additional screening of people is desirable to improve individual health outcomes, extensive screening increases the overall pressure on the healthcare system as the number of positive cases increases.
As observed in our case study, this additional workload is currently not accounted for in the resources available within the public health care system.
Despite advances in DR detection systems, the involvement of ophthalmologists remains essential~\cite{Chandakkar2012MLDR}.
In order to assess the benefit of AI-enabled CDSS, it is therefore critical that their impact is assessed not only on their image classification performance but rather their effectiveness in real-world practice.

\subsection{Limitations and Future Work}
We acknowledge several limitations in our work.
First, we merged three debiasing techniques into a single prototype to ensure a more realistic workflow.
This limits our ability to assess the suitability of individual techniques, as participants' perception was inevitably affected by their encounters with the other strategies.
Second, the bias mitigation strategies we presented can be integrated into the user interface in myriad ways.
As such, the ophthalmologists' statements regarding our strategies' implementation may not directly transfer to individual bias mitigation strategies.
However, to integrate bias mitigation strategies into CDSS, we must suggest and evaluate concrete ways of doing so.
Third, a quantitative assessment of the impact of the bias mitigation strategies was out of the scope of our study as we focused on clinicians' perceptions of bias mitigation.
While we are therefore unable to report on the impact of bias mitigation during actual real-world usage, we consider this as a critical step towards aligning with stakeholder needs.
This aligns with Bertrand et al.'s recent call for an increased emphasis on cognitive biases in the context of explainable AI~\cite{Bertrand2022CognitiveBiasReview}.

Future work may explore the effect of specific debiasing strategies and eventually study its effect in clinical trials to understand its impact in practice. 
Reflecting on our role as researchers, we chose to point to the literature when discussing cases of cognitive bias in the clinic---rather than potentially upsetting participants by pointing to their own biases.
Although several participants actively recognised the impact of their own cognitive biases on their work, not all expressed a similar sentiment.
This points to the need to increase participant awareness of potential cognitive biases, and the challenges researchers may face when involving end-users in designing cognitive bias mitigation strategies.
This study focused exclusively on ophthalmological examination, DR in particular.
Although tedious and requiring a thorough investigation, the assessment of DR is not representative of all ophthalmological tasks or clinical image assessment more broadly.
In particular, different tasks may evoke other notions of urgency, both in terms of time and clinical need.
It would therefore be valuable to study perceptions towards bias mitigation in other clinical contexts.

\section{Conclusion}
In this paper, we investigated ophthalmologists' perceptions of bias mitigation strategies as part of their AI-supported clinical workflow.
We first conducted a contextual understanding study in which we interviewed and observed domain experts to understand their use of an AI support system.
Following, we designed and evaluated an interactive prototype that embodies three bias mitigation strategies, `hear the story first', `decision justification', and `consider the opposite', as part of their existing workflow in screening diabetes patients for DR.
We find that clinicians were largely aware of the cognitive biases that may emerge when faced with AI suggestions.
However, we encountered extensive aversion towards any bias mitigation strategy that may negatively impact overall efficiency.
Based on our findings, we point to the potential of subtle design patterns to overcome aversion to bias mitigation techniques, to examine biases periodically rather than constantly, and to increase awareness of potential cognitive biases through the DSS's user interface.
Given the growing role of AI support in clinical care, we urge the HCI research community to critically assess and work to minimise any cognitive biases that may arise due to the increased reliance on AI-driven decision-support systems.

\begin{acks}
The authors would like to thank Tobias Nissen for his valuable support and express our gratitude to the clinicians who took part in the studies.
This work is supported by the `Algorithmic Explainability for Everyday Citizens' project of the Carlsberg Foundation and the `Explain-Me' project of Digital Research Centre Denmark (DIREC) under Innovation Fund Denmark.
\end{acks}

\balance

\bibliographystyle{ACM-Reference-Format}
\bibliography{references}

\end{document}